# Field evaluation of a mobile app for assisting blind and visually impaired travelers to find bus stops

*Shrinivas Pundlik [1], Prerana Shivshanker [1], Tim Traut-Savino [2], Gang Luo [1]*

1. Schepens Eye Research Institute of Mass Eye & Ear, Harvard Medical School Boston, MA
2. The Carroll Center for the Blind, MA

Corresponding Author:
Gang Luo (gang.luo@schepens.harvard.edu)

Keywords: vision aid, navigation, mobility, accessibility, deep learning, computer vision, public transportation

**Abstract**

*Purpose:* It is reported that there can be considerable gaps due to GPS inaccuracy and mapping errors if blind and visually impaired (BVI) travelers rely on digital maps to go to their desired bus stops. We evaluated the ability of a mobile app, All_Aboard, to guide BVI travelers precisely to the bus-stops.

*Methods:* The All_Aboard app detected bus-stop signs in real-time via smartphone camera using a neural network model, and provided distance coded audio feedback to help localize the detected sign. BVI individuals used the All_Aboard and Google Maps app to localize 10 bus-stop locations in Boston downtown and another 10 in a sub-urban area. For each bus stop, the subjects used the apps to navigate as close as possible to the physical bus-stop sign, starting from 30 to 50 meters away. The outcome measures were success rate and gap distance between the app-indicated location and the actual physical location of the bus stop.

*Results:* The study was conducted with 24 legally blind participants (mean age [SD]: 51[14] years; 11 (46%) Female). The success rate of the All_Aboard app (91%) was significantly higher than the Google Maps (52%, $p<0.001$). The gap distance when using the All_Aboard app was significantly lower (mean [95%CI]: 1.8 [1.2-2.3] meters) compared to the Google Maps (7 [6.5-7.5] meters; $p<0.001$).

*Conclusion:* The All_Aboard app localizes bus stops more accurately and reliably than GPS-based smartphone navigation options in real-world environments.

*Translational Relevance:* The All_Aboard mobile app navigation aid can be potentially help BVI individuals independently travel via public transportation.

**Introduction**

Blind and visually impaired (BVI) people often rely on public transportation, such as buses and subways, to travel for employment, leisure or for other needs.[1] Geolocation and transportation information accessed through smartphones has greatly facilitated macro-navigation for BVI people. For example, one can plan a route and get detailed instructions on mobile devices for point-to-point navigation using public transits. Navigation apps make up one of the major groups of vision assistance mobile apps available in App store and Play store. [2] On the other hand, micro-navigation – navigating precisely to the desired destination at any stage of the journey, remains a largely unsolved issue for BVI individuals. To comply with the Americans with Disabilities Act (1990), regional transit agencies are required to comply with regulations regarding the accessibility of transit infrastructures.[3] In the context of vision disabilities, the requirements include placing of large-print signage at bus stops, providing braille and tactile information within transit stations, and making stop announcements inside transit vehicles at main points, among others. However, poor interface and lack of cues accessible from distance, for example transit stop signage, are one of the main barriers to equal access to public transportation.[4-9]

Systemic inaccuracies in the GPS based location services is the underlying problem that leads to micro-navigation challenges faced by people with BVI. This is also referred to as the last 30 feet or last 10 m problem in wayfinding. For example, when navigating to a bus stop, a blind person following any GPS-based navigation app, may arrive at the app indicated location that has a considerable gap (typically 10 m or 30 feet) from the actual bus stop due to the localization error in the GPS service. For perspective, this 10 m gap could be almost equal to an entire bus length in certain regions. According to the feedback from blind travelers, sometimes even a small gap can be large enough for them to miss the bus because the bus drivers misunderstand their intention and not stop for them.[10-13] In the worst-case scenario, especially in crowded cities, the GPS localization may be off by more than a block, making the macro navigation apps essentially useless in pedestrian mode.[14]

Weather and location (density of tall buildings in downtown areas for example) can further affect GPS-based localization. In addition to localization error, there is a possibility of mapping errors (sometimes large) in the stop locations made publicly available by the transit agencies. In our survey of 174 bus stop locations in Boston metro area, about 23% were mapped more than 2 bus lengths away.[15] Mapping and localization errors together contribute towards making purely location-based services unreliable for micro-navigation tasks, such as finding bus stops. Making matters worse, bus stop signs can be one of many signs on a typical urban street (among traffic/parking signs and street signs), and thus finding it becomes a visual search task, in addition to a plain geolocation task. Since visual search performance is known to be significantly degraded in people with low vision,[16,17] it is evident that a navigation aid is needed for micro-navigation with visual search capabilities.

One of the conventional wayfinding solutions is the use of Bluetooth beacons to provide micro-location information with high accuracy on nearby landmarks.[18-22] This approach's scalability and applicability in outdoor environments are restricted due to the high cost for infrastructure modification and maintenance. On the other hand, smartphones could allow rapid scaling of accessibility. A few smartphone apps have been developed, tested, or released to help blind and visually impaired people access public transportation specifically, or to navigate to destinations in general.[23-25] These apps are primarily GPS-based, and therefore still subject to the limitations of GPS-based navigation systems detailed above. In order to achieve localization more accurately, some apps combine location information together with landmark recognition.[12,13,26,27] However, landmark maps around the various locations have to be built, maintained, and made widely available prior to use. Combing signage information extracted from GTFS with optical character recognition could provide a viable micro-navigation solution.[28]

For the purpose of bus stop navigation, a purely visual approach can work well if combined with a typical macro-navigation apps.

We have developed a mobile app, All_Aboard that recognizes bus stop signs to help the users navigate within a very short range of the physical location of the sign.[29] When using the app, users can first use commonly available macro-navigation tools, such as Google Maps, to arrive within the vicinity of the bus stop location, and then they can scan the surroundings with their smartphone camera. The All_Aboard app detects the bus stop signs in real-time in the smartphone camera imagery, and can guide the users to approach bus stops through auditory cues, with the auditory pitch coding the distance to the target. Our preliminary testing of the All_Aboard app indicated its superior performance compared to Google Maps app.[14]

The goal of this study was to evaluate All_Aboard in real-world conditions with BVI transit users and compare its localization ability with navigation via Google Maps app. Our primary hypothesis was that the localization based on All_Aboard app was significantly better than just using a conventional navigation app (Google Maps) in terms of distance to the desired bus-stop location and rate of successful localizations. Given that GPS-based localization typically suffers in densely built downtown areas, we further hypothesized that All_Aboard might be more effective in these locations compared to more sparsely populated sub-urban areas.

**Methods**

*All_Aboard App*

One of the underlying ideas behind the All Aboard app is that the bus stop signage is unique (different than other road signs), uniform in appearance (typically, but not always), and standardized across the entire area of a transit agency (Figure 1A). Moreover, since the bus stop signs for a given transit agency have the same known physical size, one can estimate the approximate distance based on the image size (width) of the detected bus stop sign (the farther the distance, the smaller the image, and vice versa). Therefore, it is feasible for computer vision algorithms to learn the appearance of the bus stop signs, detect them in the images captured by smartphone cameras, and estimate the distance to the actual sign. Bus stop detection in All_Aboard is performed in real-time using a MobileNetv2 deep learning neural network,[30] trained on about 10000 images of bus stops collected for a given city/region. Images of bus stop signs were collected from the Google Street View imagery based on the publicly available bus stop coordinates via General Transit Feed Specification (GTFS) standard. The stop signs were manually labeled by placing a bounding box in the collected images. The trained model runs natively on the smartphone device (no cloud processing).

Another key idea behind the operation of All_Aboard is that it supplements a macro-navigation app and thus only need to be operational in the general vicinity of the bus stop. In a typical usage scenario (Figure 1B), the user launches All_Aboard when a macro-navigation app (Google Maps etc.) indicates that the user is near the desired bus stop location. The All_Aboard app senses phone orientation and searches only when the device is held in an upright position. The user can then scan with the phone camera in an arc to first determine the angular orientation of the sign. A positive detection leads to a beeping sound and a true positive detection is typically indicated by a series of continuous auditory tones. Once locked in the direction of the sign, the auditory tones change in frequency as the user get near the bus stop sign (similar to a homing signal). Thus, the app can help users gauge relative distance to the sign and adjust their approach. There were four levels of audio tones, with the highest frequency indicated that the detected sign was within 2 meters (≈ 6 feet). After it is launched, the apps works without needing any active intervention from the user, as long as the device is held in an upright manner.

The All_Aboard is available in App store and Play store for free, and it is capable of recognizing bus stops in 10 major cities/regions around the world. Once installed, the users can download the available trained neural network models for the corresponding transit agencies. In this study, the All_Aboard app was loaded with the model trained to detect bus stop signs of the Massachusetts Bay Transportation Authority (MBTA) that operates public transit in Boston metro area.

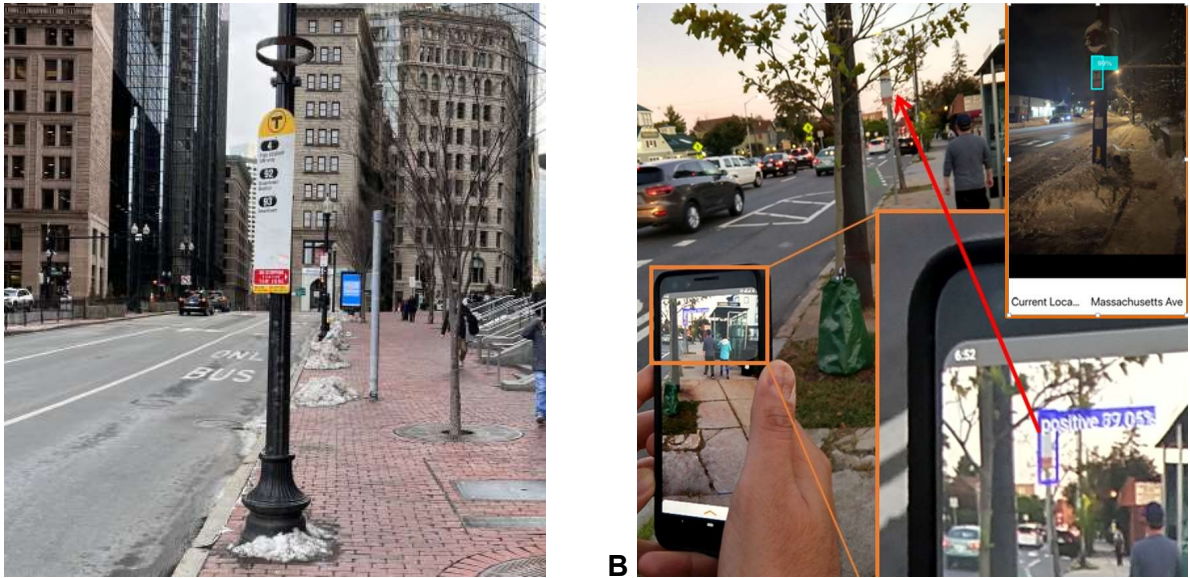

**Figure 1:** (A) A typical MBTA bus stop sign at one of the trial location in downtown Boston. This is the most recent version of the signage, however, older versions with slightly different appearance but similar shapes persist throughout the coverage area. Very few bus stops in the area covered by the transit agency are sheltered, and the distinctive sign is often the only visual identification of the bus stop. (B) Operation of the All_Aboard app in the general vicinity of a bus stop. The lower inset shows the app detecting the bus stop sign (the bounding box is drawn around the detected sign in the camera view displayed on the smartphone screen. The upper inset shows a successful detection at night in low light conditions. A demonstration video of All_Aboard app in action can be found at https://www.youtube.com/watch?v=VUVpqEw1_2k.

*Participants*

The inclusion criteria were: vision status of legal blindness, independent mobility without assistance from a sighted guide, physical ability to walk over a distance of about 1 mile at a given time, and familiarity with smartphone/mobile devices. Participants for this study was recruited via referrals from the Carroll Center for the Blind Newton Massachusetts, practitioners at vision rehabilitation clinics, and via a pool of volunteers who had participated in prior studies. The study protocol was approved by the Institutional Review Board at Mass Eye & Ear. The study followed the tenets of the Declaration of Helsinki and written informed consent was obtained from all the participants. The participants were reimbursed for travel to the study sites and for their time.

*Study Design*

The study involved 2 visits at 2 separate study sites: downtown Boston (City) and near the campus of the Carroll Center for the Blind in Newton, Massachusetts (Suburb). Each study site involved navigating to 10 bus stops following a specific route (Figure 2). For each bus stop, performance with All_Aboard and Google Maps apps was evaluated with both the apps running simultaneously. During the study, the participants were accompanied by a certified orientation and mobility instructor (O&M) who provided directions along the route and ensured the safety of the participants during the study. Members of the study team also accompanied the participant and the O&M instructor, who administered the study and recorded measurements. For all trials, a preconfigured Android smartphone was provided to the participants.

Before starting the study, each participant was provided oral instructions and hands-on training with using the All_Aboard app at a practice location. At the start of the trial at each bus stop location, the participant was guided by O&M instructor to a location that was about 30 to 50 meters (≈100 to 150 feet) away from the bus stop sign along the direction of travel in approximately straight ahead direction. The starting distance from the stop sign was varied at each stop location to dissuade participants from guessing the stop location based on step counting. The path from the starting location to bus stop sign did not involve crossing streets, except in the case of one bus stop location in Newton, where the stop sign was affixed very close to the intersection. At the starting point for each trial, Google Maps app (operating in the pedestrian directions mode) was launched by the experimenter, StreetView calibration was done (this was one of the features of the Google Maps app that uses live camera imagery to geo-locate more accurately), and the mapped location of the said bus stop was set as the destination. Then, the All_Aboard app was launched such that both apps were running simultaneously, with Google Maps navigation window in the inset at the bottom of the screen (Figure 3A).

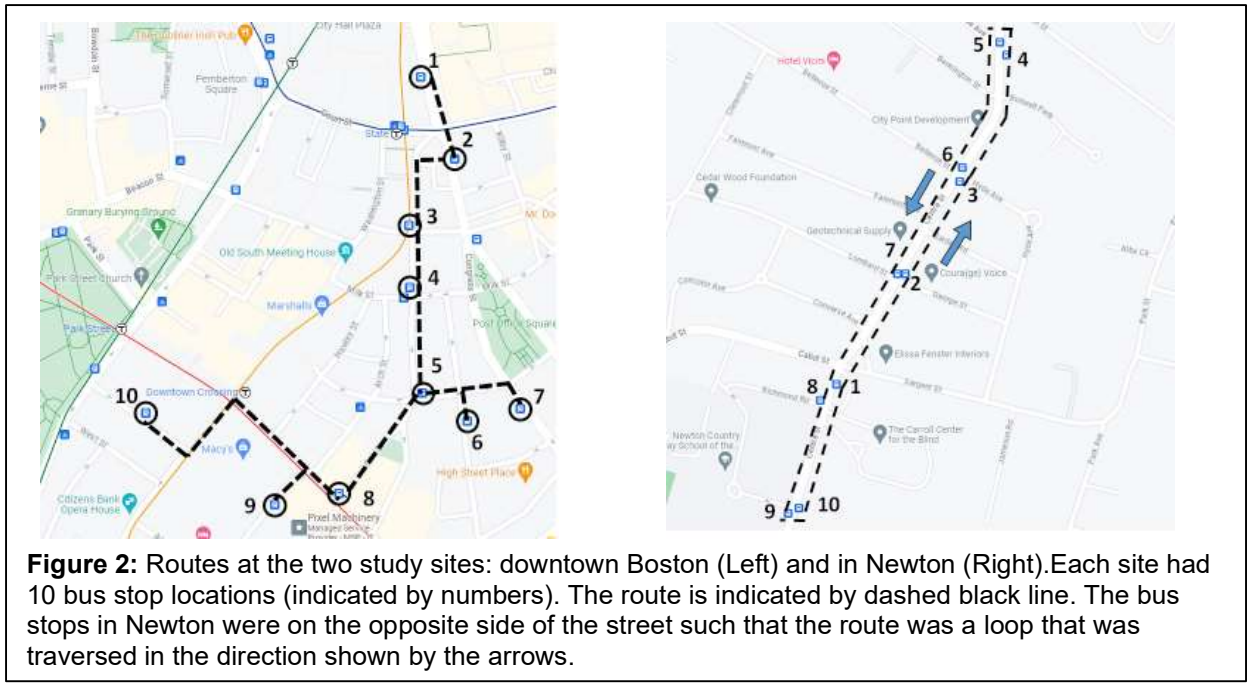

**Figure 2:** Routes at the two study sites: downtown Boston (Left) and in Newton (Right). Each site had 10 bus stop locations (indicated by numbers). The route is indicated by dashed black line. The bus stops in Newton were on the opposite side of the street such that the route was a loop that was traversed in the direction shown by the arrows.

Then, the smartphone device was handed over to the participants. From this starting location, the participants was instructed to navigate as close to the bus stop sign as possible. They were also instructed to hold the smartphone upright with its rear camera pointing straight ahead (Figure 3B), and scan side-to-side to determine the relative orientation of the bus stop sign from their walking trajectory. After this point, the participants walked on their own, relying on their habitual mobility aid and their residual vision if present, along with the auditory feedback from the All_Aboard app. Meanwhile, the Google Maps app provided intermittent voice instructions, including distance to the destination in feet (which the participants soon learned were often unreliable).

At the end of the trial at a bus stop location, the participants stopped and notified the experimenters when they thought they were closest to the bus stop sign as per their judgment. This was primarily based on the audio feedback by the All_Aboard app – when the audio tone frequency and pitch were at the highest levels indicating the detected sign was very close. A few participants could use their residual vision from this point onward to get even closer. Distance from where they stopped to the actual stop sign was measured with a tape measure. This was the localization distance for All_Aboard app. Google Maps app also indicated via auditory feedback when it determined that the participant arrived at the destination (Figure 3C). The distance between the bus stop sign and the arrival point according to Google Maps was also measured with the tape measure.

At the time of their first study visit, the participants answered a brief survey that collected basic demographic information, vision status, use of vision aids, and their preferred transit options (public transit, rideshare, or private vehicle –family member driving). The level of vision was recorded either in terms of self-reported visual acuity in Snellen, or as light perception, or no light perception (in case of completely blind individuals).

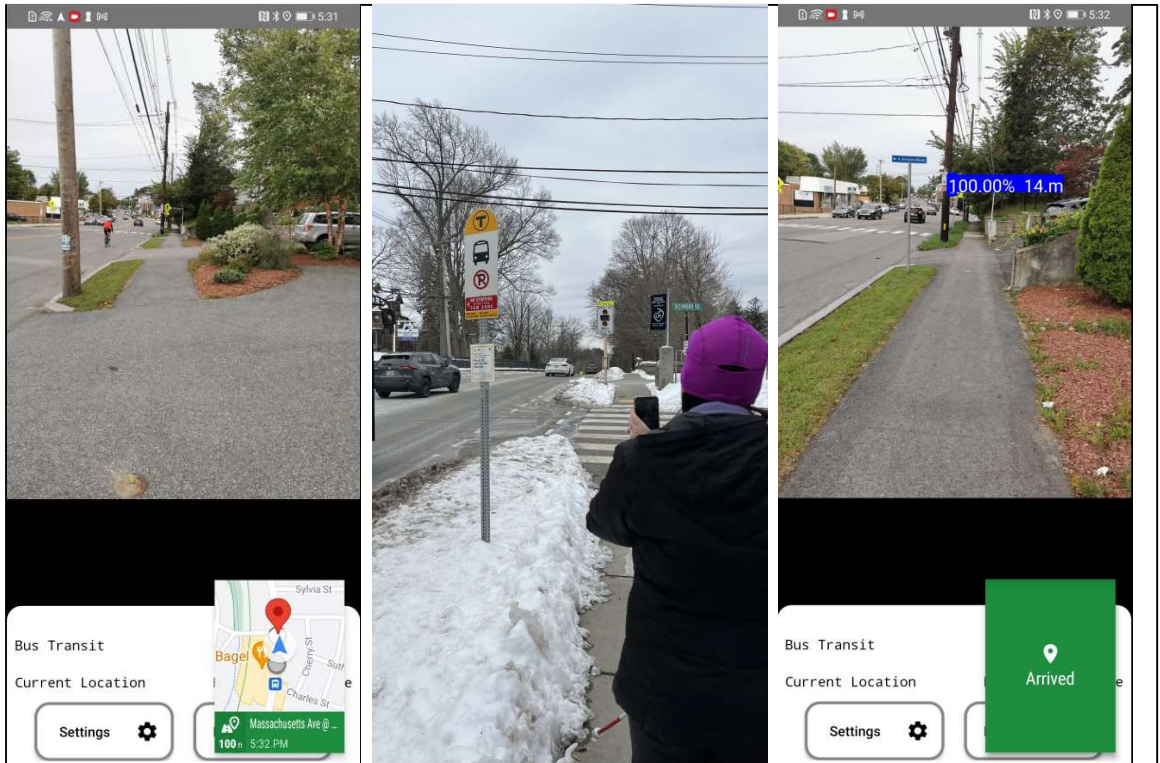

**Figure 3:** Running All_Aboard and Google Maps simultaneously to find bus stops. (A) Screenshot of the device at the start point. Both All_Aboard and Google Maps (inset) launched and run simultaneously. (B) An user holds the phone upright with the rear camera facing straight ahead. (C) Screenshot of the device when Google Maps indicate arrival at the destination. The All_Aboard app indicates the physical bus stop sign is still some distance ahead.

*Outcome Measures*

The two main outcome measures, separately obtained for each app (All_Aboard and Google Maps), were: the localization error (gap distance) in meters and the rate of successful localizations (success rate). As mentioned above, the gap distance was obtained via direct measurement of the distance between app indicated/subject determined location of the bus stop and the physical location of the bus stop sign. When the All_Aboard guided the participant close enough for them to spot (via their residual vision) or identify the bus stop sign or touch the pole or post, the gap distance was marked as 0. If the app indicated location on the ground was beyond the physical bus stop sign with respect to the direction of travel, then the measured gap distance was recorded as negative. A trial instance was deemed as a failure if a reasonable measurable distance was not obtained. Success rate for each app was defined as % of locations for which a valid measurable distance along the travel path was available.

At any given bus stop location, trial failures could occur because of various reasons. In case of Google Maps app, incorrect mapping of the bus stops was one of the reasons. Such failures were predictable and repetitive across all the subjects because the location of the bus stop in the map was fundamentally incorrect. One cause of trial failure was the mapped location being more than 100 feet away from the physical bus stop. Another cause of trial failure with Google Maps app was catastrophic inaccuracies in geolocation and consequent navigation directions, for example, when the app directed the users to cross streets, double-back, or go around a corner, which would lead them completely miss the bus stop. These failures were more prevalent in the downtown Boston area with tall buildings and/or on overcast/rainy days.

In the case of All_Aboard app, trial failure could occur because of detection failures due to false negatives or deficient technique by the subjects while using the app. Shadows and occlusions could lead to the app to fail to recognize a bus stop sign. Signage largely slanting away from the side walk direction can cause the app fail to detect. On other occasions, the subjects did not scan sufficiently while walking towards the bus stop sign and lost the audio signal. If the bus stop sign was initially detected but then went outside the field of view of the camera as the subject approached, the continuous audio signal suddenly stopped. This was an indicator to the subjects that they either passed the sign or are too close to it. They were allowed to retrace their steps and try again once to zero-in or confirm the presence of the stop sign in the near vicinity. If major intervention by O&M or the experimenter was needed to reorient the subject after initial failure to detect, then the trail was considered as a failure for the All_Aboard app for the given location, even if the sign was successfully detected in the subsequent tries.

*Statistical Analysis*

Potential factors of interest affecting the outcome measures were app used (All_Aboard or Google Maps), the study location (City vs. Suburb), and the vision status (with or without residual vision). Completely blind subjects without light perception were categorized as without residual vision, while the rest were with residual vision. Vision in the better eye was used for this categorization. Association of gap distance with these above potential variables was analyzed within-subject via a linear model in repeated measures framework. The success rate was analyzed using a binary logistic regression. In addition to the main effects, the interaction between the above 3 factors were also examined. Estimated marginal means with their 95% confidence intervals and contrasts are reported for gap distance. Estimated mean marginal probability of success and the 95% confidence interval is reported from the logistic regression model for success rate. *P* values <0.05 were considered statistically significant. Statistical analysis was performed using statistical packages in R (ver. 4.0.4).[31-36]

**Results**

Total 25 subjects were recruited, of which, data for both study sites was available for 24 subjects (see Table 1 for summary of subject characteristics). One subject was dropped from the trial after first visit due to the concerns about overall physical fitness required to complete the study. Eleven participants (46%) were female. A variety of conditions affected the vision of the participants. While all were legally blind in the US, their vision ranged from completely blind to 20/200 vision. The majority walked with long cane and all except one used an iPhone in their daily lives. Public transit was the most preferred transit option, followed by rideshare and private vehicle.

**Table 1:** Study participant characteristics.

| | |
|---|---|
| N | 24 |
| Age (years) | Median: 55, IQR: 41 – 61, Min.: 20, Max.: 71 |
| Gender | Female: 11 (46%) |
| Vision | Available VA measure – N: 11 (46%), range: [20/200 – 20/1200]<br>Light Perception – N: 6 (25%)<br>No Light Perception – N: 7 (29%) |
| Vision disorders | RoP: 5, retinitis pigmentosa: 4, retinal detachment: 3, glaucoma: 2;<br>One case each of: age-related macular degeneration, optic atrophy, aniridia, cone dystrophy, retinal artery occlusion, charge syndrome, monochromacy, diabetic retinopathy, Norrie syndrome |
| Duration of vision loss | At birth: 14 (58%)<br>Acquired: 8, Median duration: 18 years, range: [3 – 41 years]<br>Data not available: 2 |
| Mobility aids used | Long cane: 16 (67%)<br>Guide dog: 6 (25%)<br>None: 2 (8%) |
| Most preferred transit option | Public Transit - 1st: 11, 2nd: 7, 3rd: 3, NA: 3<br>Rideshare - 1st: 9, 2nd: 11, 3rd: 2, NA: 2<br>Private car - 1st: 4, 2nd: 5, 3rd: 8, NA: 7 |
| Habitual smartphone | iPhone 23 (96%) |

Table 2 shows the trial instances and other data for both apps and at each study site. Across 24 subjects, there were supposed to be trials at 480 designated bus stops. However, over the course of the study, some bus stops were skipped due to construction or missing bus stop signs, resulting in a total of 48 instances with missing data. Therefore, each app was evaluated in a total of 432 instances. Overall success rate and gap distance measures were substantially better with All_Aboard app than Google Maps.

**Table 2:** Cumulative statistics for trial data.

| | | Both Apps | Google Maps | All_Aboard |
|---|---|---|---|---|
| Bus stop instances with available data | Both sites | 864 | 432 | 432 |
| | City | 458 | 229 | 229 |
| | Suburb | 406 | 203 | 203 |
| Skipped instances | Both sites | 96 | 48 | 48 |
| | City | 22 | 11 | 11 |
| (missing data) | Suburb | 74 | 37 | 37 |
| Number of successful instances | Both sites | 626 | 225 | 401 |
| | City | 329 | 112 | 217 |
| | Suburb | 297 | 113 | 184 |
| Success rate (%) | Both sites | 72 | 52 | 93 |
| | City | 72 | 49 | 95 |
| | Suburb | 73 | 56 | 91 |
| Average [SD] gap distance (m) | Both sites | 3.36 [3.65] | 6.62 [4.15] | 1.54 [1.36] |
| | City | 3.04 [3.82] | 6.26 [4.89] | 1.38 [1.34] |
| | Suburb | 3.72 [3.41] | 6.97 [3.24] | 1.72 [1.36] |

In Table 2, successful instances with each app are listed independently of each other. When compared pairwise at each bus stop instance (Table 3), there were only a handful of instances where both apps failed (18 out of 432 or about 4%). There were 13 (3%) instances where All_Aboard failed but Google Maps succeeded, and there were 189 (44%) instances where Google Maps failed but All_Aboard succeeded. For the former cases, the average [SD] gap distance was 9.3[5] meters, and for the latter cases, the average [SD] gap distance with All_Aboard was 1.6[1.4] meters. From 225 successful instances with Google Maps, the arrival location was mapped past the bus stop sign along the direction of travel in 60 instances (about 27%), with an average gap distance of 7.2[2.9] meters.

**Table 3:** 2x2 tables showing joint successes or failures of the All_Aboard and Google Maps over all 432 bus stop instances.

| | Overall | | City | | Suburb | |
|---|---|---|---|---|---|---|
| | Google Maps Success | Google Maps Failure | Google Maps Success | Google Maps Failure | Google Maps Success | Google Maps Failure |
| All_Aboard Success | 212 | 189 | 106 | 116 | 106 | 78 |
| All_Aboard Failure | 13 | 18 | 6 | 6 | 7 | 12 |

There was no significant effect of subject age, gender, or the kind of mobility aid used on the gap distance or on the success rate. The results and discussion is mostly related to 3 key factors: the app used, study site, and subject group based on their vision status.

Gap distance (in meters), averaged over vision status and study sites, was significantly smaller with All_Aboard (mean: 1.8, 95% CI: 1.3-2.3) compared to Google Maps (mean: 7.0, 95% CI: 6.5-7.5; p<0.001). Gap distance with All_Aboard was significantly smaller than Google Maps in City and Suburb, as well as in subjects with or without residual vision (Figure 4A). The gap distance was significantly larger in completely blind group (mean: 8.4, 95% CI: 7.3-9.5) compared to those with residual vision (mean: 5.4, 95% CI: 4.7-6.1; p<0.001) in the City with Google Maps. No significant effect of vision status on the gap distance with All_Aboard was observed. A significant effect of study site was seen only in the case of Google Maps in subjects with residual vision, where gap distance in the Suburb (mean: 6.8, 95% CI: 6.1-7.5) was significantly larger than that observed in the City (mean: 5.4, 95% CI: 4.7-6.1; p<0.022). Again, no significant effect of study site was observed for gap distance resulting from the All_Aboard app.

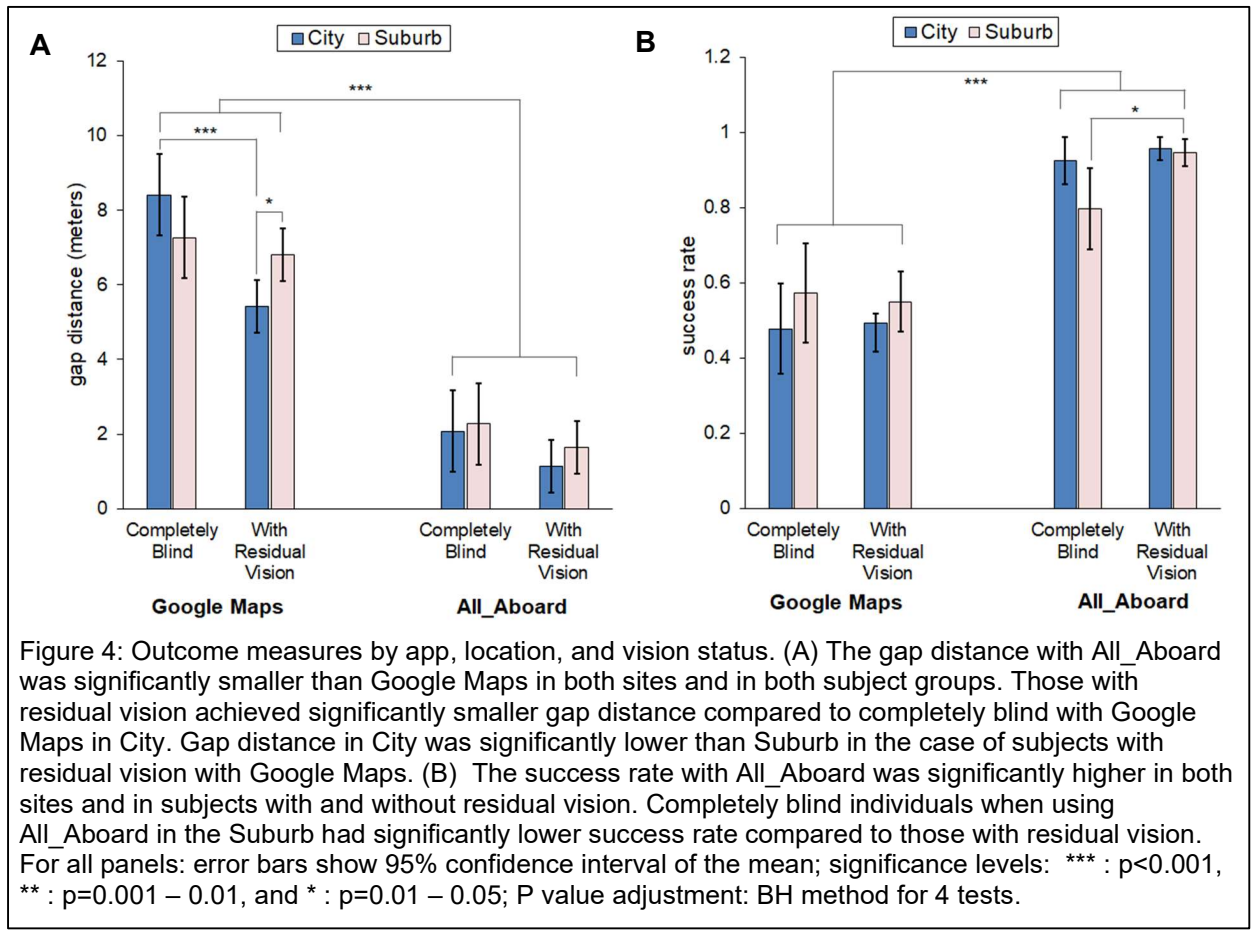

Figure 4: Outcome measures by app, location, and vision status. (A) The gap distance with All_Aboard was significantly smaller than Google Maps in both sites and in both subject groups. Those with residual vision achieved significantly smaller gap distance compared to completely blind with Google Maps in City. Gap distance in City was significantly lower than Suburb in the case of subjects with residual vision with Google Maps. (B) The success rate with All_Aboard was significantly higher in both sites and in subjects with and without residual vision. Completely blind individuals when using All_Aboard in the Suburb had significantly lower success rate compared to those with residual vision. For all panels: error bars show 95% confidence interval of the mean; significance levels: *** : p<0.001, ** : p=0.001 – 0.01, and * : p=0.01 – 0.05; P value adjustment: BH method for 4 tests.

The success rate with All_Aboard was significantly higher than Google Maps across both study sites and both the subject groups (Figure 4B). The overall success rate with All_Aboard (mean: 0.91, 95% CI: 0.87-0.94), averaged over study sites and subject group factors, was about 75% higher than Google Maps (mean: 0.52, 95% CI: 0.47-0.58; p<0.001). When using All_Aboard in the Suburban location, the success rate for completely blind subjects (mean: 0.8, 95% CI: 0.69-0.90; p<0.001) was slightly but statistically significantly lower compared to those with residual vision (mean: 0.95, 95% CI: 0.91-0.98; p<0.001). Otherwise, there was no significant difference in success rates for any other conditions.

**Discussion**

When people (normally sighted or BVI) take buses in areas like Boston metro region, where most bus stops are indicated just by a sign on a post, standing even a short distance away from the sign may cause the buses do not stop for them. This accessibility challenge may diminish independence, compromise adoption of affordable transportation for BVI travelers.[1] This is just one of the last-10-meter navigation assistance needs of BVI individuals that is unmet. In this study, we evaluated the ability of All_Aboard app relative to the Google Maps navigation app in guiding BVI travelers accurately to bus stop locations in urban and suburban settings. The rate of successful localization of bus stops was substantially higher and the gap distance was much smaller when using All_Aboard app compared to Google Maps navigation. On average, the All_Aboard app was able to guide the subjects within about 2 meters (6 feet) of the bus stop sign, whereas with Google Maps they were likely to be about 7 meters (23 feet) away. The large effect size of All_Aboard app in terms of success rate of localization and the gap distance was observed irrespective of the location of testing, the vision status of the subjects, other demographic characteristics, and the kind of mobility aids they used. Thus, our findings demonstrate that All_Aboard app could provide a reliable benefit in navigation by accurately detecting the bus stop sings and guiding the users close enough to the designated stop that makes it less likely that a bus will pass by them for standing too far from the bus stops. Importantly, this study validates that computer vision-based object recognition capabilities can be used in a complementary way and provide added benefit to purely location-based navigation services in real-world settings.

During the study design, we were expecting some difference in performance in a city vs. suburban location based on our previous preliminary study,[14] due to the well-known limitations of location-based services in areas with tall structures. Therefore, we did not expect a significant effect of location on All_Aboard app, and the findings were more or less consistent with this expectation. In case of Google Maps, City vs. Suburb setting did not have any significant effect on the success rate, and its effect on gap distance relatively modest (slight but statistically significant difference was seen only in subjects with residual vision). The primary reason for this lack of a location effect was that the localization error in the City was somewhat balanced or counteracted by large mapping errors in the Suburb. Thus, despite better localization accuracy of Google Maps in the Suburb, large mapping errors meant that almost the same proportion of trials were unsuccessful as in the City.

While we enrolled participants with a wide range of visual abilities – from completely blind up to visual acuity of 20/200, we analyzed only the effect residual vision presence on the performance with the navigation apps. If the All_Aboard app could successfully guide low vision travelers close to the bus stop sign, it was possible that they could use their residual vision (even if it was only restricted to shape or perform perception) from there on to navigate further close to the sign. We indeed observed this behavior in a few participants. However, as a whole, gap distance was not significantly different between those with or without residual vision – indicating that the app already guided the subjects close enough to the bus stop sign, such that any further change due to residual vision was not large in terms of distance. However, trial success rate was affected by residual vision presence in the Suburb, as completely blind subjects experienced significantly more failures with All_Aboard app compared to those with residual vision. A possible reason for higher failure rate in the Suburb could be because some of the bus stop signs on that site were not properly placed. Some were occluded by trees, slanting towards the street instead of the sidewalk, or not at the edge of street curb. In these situations, scanning sufficiently wide is crucial. However, completely blind subjects tended to scan across a narrow range or not scan at all due to complete loss of visual input to help with orientation. Therefore, they were more likely to miss signs that are slightly more difficult to find. More training and practicing on scanning skills might help improve their success rate in these situations.

**Acknowledgments**

The All_aboard app development was funded in part by Microsoft AI4A award. The authors would like thank Nick Corbett and Dinna Rosenbaum from the Carroll Center for the Blind for their help with participant recruitment and coordination.